\documentclass[12pt]{article}

\catcode`\@=11

\global\arraycolsep=2pt
\usepackage{amsbsy,amssymb,latexsym,amsfonts,amsmath}
\usepackage{graphicx,color}

\allowdisplaybreaks

\begin{document}
\begin{flushright}
\parbox{4.2cm}
{RUP-20-13}
\end{flushright}

\vspace*{0.7cm}

\begin{center}
{ \Large Bootstrap bound on extremal Reissner-Nordstr\"om black hole in AdS}
\vspace*{1.5cm}\\
{Yu Nakayama}
\end{center}
\vspace*{1.0cm}
\begin{center}

Department of Physics, Rikkyo University, Toshima, Tokyo 171-8501, Japan

\vspace{3.8cm}
\end{center}

\begin{abstract}
Based on the numerical conformal bootstrap bound, we show that the arbitrarily small Reissner-Nordstr\"om black hole in AdS space-time is inconsistent with holography unless the energy spectrum is modified quantum mechanically or it is unstable as indicated by the weak gravity conjecture.
 
\end{abstract}

\thispagestyle{empty} 

\setcounter{page}{0}

\newpage

\section{Introduction}
The Reissner-Nordstr\"om black hole in AdS space-time is a ubiquitous solution of various effective field theories of gravity, yet it has many puzzling issues. First of all, unlike in asymptotically Minkowski space-time, it does not saturate the BPS bound even if it is extremal or at zero temperature. It could suggest that the extremal Reissner-Nordstr\"om solution in AdS space-time is unstable, but without extra charged matter, its classical instability has not been explicitly shown \cite{Gubser:2000mm}\cite{Konoplya:2008rq}\cite{Gwak:2015ysa}. On the other hand, the non-trivial energy spectrum with respect to the charge coincides with the large charge universality of $U(1)$ symmetric conformal field theories \cite{Hellerman:2015nra}\cite{Nakayama:2015hga}\cite{Loukas:2018zjh}. We then expect a non-trivial role of the extremal Reissner-Nordstr\"orm black hole in holographic conformal field theories. 

The (in)stability of the extremal Reissner-Nordstr\"om black hole has attracted a lot of attention in relation to the ``weak gravity conjecture"\cite{ArkaniHamed:2006dz} (see e.g. \cite{Palti:2019pca} for a review). Conceptually, the weak gravity conjecture claims that the gravity must be weaker than the electromagnetic force so that the (extremal) Reissner-Nordstr\"om black hole must decay. Beyond the classical limit in the Minkowski space-time, however, the claim becomes ambiguous, and making the conjecture precise has been of theoretical interest to understand the nature of quantum gravity better. Through holography, it may also imply a non-trivial constraint on conformal field theories \cite{Nakayama:2015hga}.

Furthermore, as pointed out in \cite{Crisford:2017gsb}\cite{Horowitz:2019eum} there is a mysterious connection between the weak gravity conjecture in the AdS space-time and a possible appearance of a naked singularity in the same theory. The formation of the naked singularity can be avoided by the effect of a light charged scalar field. At the same time, the existence of such a light charged scalar field indicates that the extremal Reissner-Nordstr\"om black hole is unstable due to the superradiance. The instability caused by the superradiance leads to a condensation of the charged scalar field and in certain supersymmetric situations, we may end up with a hairy black hole that saturates the BPS bound, which becomes the lightest charged state in the AdS space-time.

We would like to understand these peculiar features of the extremal Reissner-Nordstr\"om black hole in the AdS space-time from holography. 
For this purpose, in this paper we will discuss the numerical conformal bootstrap bound on the $U(1)$ charged object to make a non-perturbative statement. Before doing any numerical study, one can immediately predict what should happen to the energy level of the (would-be) extremal Reissner-Nordstr\"om black hole at the quantum level.
Suppose that the minimal charged extremal Reissner-Nordstr\"om black hole saturates the unitarity bound of the AdS energy (i.e. $\Delta = \frac{D-3}{2}$ in $D=d+1$ space-time dimensions) then the charge two extremal Reissner-Nordstro\"om black hole has the AdS energy $\Delta_2 > 2\Delta$ classically, but this cannot be the case: the existence of a dual conformal field theory demands $\Delta_2 = 2\Delta$. In the main part of the paper, we will make stronger quantitative constraints from the numerical conformal bootstrap analysis.

There are several scenarios to avoid the inconsistency from the gravity side. One possibility is that the extremal Reissner-Nordstr\"om black hole remains stable, but the quantum correction makes the energy spectrum modified so that the conformal bootstrap bound is satisfied. The other possibility is that the weak gravity conjecture holds: there always exist states with lower energy than the extremal  Reissner-Nordstr\"om black hole and these states, possibly non-gravitational objects or hairy black holes, will saturate the conformal bootstrap bound. In either way, our bound will give a criterion when something more than the Einstein gravity coupled with Maxwell field should emerge.

\section{Energy spectrum of Reissner-Nordstr\"om black hole in AdS}
Let us consider the Einstein-Maxwell system with the cosmological constant in $1+3$ dimensional space-time. The  classical action  is given by
\begin{align}
S = \int d^4x \sqrt{-g}  \frac{1}{2\kappa^2} \left( R + \frac{6}{L^2}  -\frac{1}{4e^2}F_{\mu\nu}F^{\mu\nu} \right) \ .
\end{align}
It has a classical solution of the AdS space-time with radius $L$, and we will work in the asymptotic AdS space-time in the global coordinate.

As the simplest charged black hole solution of the classical equations of motion, the Reissner-Nordstr\"om-AdS metric is given by 
\begin{align}
ds^2 = -f(r) dt^2 + f^{-1} (r) dr^2 + r^2 d\Omega_2^2 \ ,
\end{align}
where $d\Omega_2^2$ is the metric of a unit two-sphere, and 
\begin{align}
f(r) = 1 - \frac{2M}{r} + \frac{ Q^2}{4r^2} + \frac{r^2}{L^2} \ ,
\end{align}
where $M$ and $Q$ are related to mass and charge of the black hole respectively. From this expression, one can read the location of the outer horizon $r= r_+$ as a larger solution of $f(r_+)=0$.

The black hole solution is supported by the gauge potential (associated with the field strength $F_{\mu\nu} = \partial_\mu A_\nu - \partial_\nu A_\mu$) given by\footnote{While we focus on the electric black hole, the following discussions also apply to magnetic or dyonic black holes.}
\begin{align}
A_t = \frac{e Q}{r} \ . 
\end{align}
We define the extremal limit by setting 
\begin{align}
 Q^2 = 4 r_+^2\left(1+\frac{3}{L^2}r_+^2\right) \ , \label{extremal}
\end{align}
under which $f(r) = 0$ has a double zero and the temperature of the black hole becomes zero. When charge $Q$ is larger than this extremal value with a fixed mass parameter $M$, the naked singularity appears in the solution.
In order to obtain the most non-trivial constraint, we always assume that the black hole is extremal in the following. By eliminating $Q$ from the extremal condition \eqref{extremal}, the mass of the extremal Reissner-Nordstr\"om black hole as a function of the horizon radius $r_+$ becomes
\begin{align}
M = r_+ \left(1+\frac{2 r_+^2}{L^2} \right) \ . 
\end{align}

For our purpose, we now compute the AdS energy of the extremal Reissner-Nordstr\"om black hole as a function of the dimensionless charge $q = \frac{ Q}{2L}$: 
\begin{align}
\Delta(q) = \frac{8\pi LM}{\kappa^2} &= c \frac{\sqrt{-1+\sqrt{1+12q^2}}(2+\sqrt{1+12q^2})}{3\sqrt{6}} 
\end{align}
where $c= \frac{8\pi L^2}{\kappa^2}$ is the ``central charge" of the dual conformal field theory.  The dimensionless AdS energy $\Delta(q)$ will be identified with the conformal dimension under the AdS/CFT correspondence. In the small charge limit, it is expanded as
\begin{align}
 \Delta(q) =c\left( q + \frac{1}{2} q^3 -\frac{9}{8}q^5 + \frac{81}{16} q^7 -\frac{3861}{128} q^9 + \cdots \right) \label{q} 
\end{align}
while in the large charge limit, it is expanded as
\begin{align}
 \Delta(q) = c\left( \frac{2}{3^{4/3}} q^{3/2} + \frac{1}{2 \cdot 3^{1/4}} q^{1/2} -\frac{1}{16  \cdot 3^{3/4}} q^{-1/2} + \frac{1}{576 \cdot 3^{1/4}} q^{-3/2} + \cdots \right) \ . 
\end{align}
As observed in \cite{Loukas:2018zjh}, this expression remarkably coincides with  a prediction from the lowest derivative effective theory of $U(1)$ symmetric conformal field theory on $\mathbb{S}_2 \times \mathbb{R}$ in the large charge limit.

Given an extremal Reissner-Nordstr\"om black hole with charge $q$, the AdS energy of the twice charged extremal Reissner-Nordstr\"om black hole is given by 
\begin{align}
\Delta_2 = \Delta(2q) = c \frac{\sqrt{-1+\sqrt{1+48q^2}}(2+\sqrt{1+48q^2})}{3\sqrt{6}} \ .  \label{2q}
\end{align}
It is important to realize that it is strictly larger than $2\Delta(q)$, implying that the ``bound state energy" is always positive even for the extremal black hole. It also means that the extremal Reissner-Nordstr\"om black hole does not saturate the BPS condition: $\Delta(q) = cq$ (or $M = \frac{1}{2}Q$). This is the distinct feature of the extremal Reissner-Nordstr\"om black hole in the AdS space-time in contrast to the one in the Minkowski space-time.

In the next section, we will study the conformal bootstrap bound. It gives a bound on $\Delta_2$ as a function of $\Delta(q)$, so we may want to eliminate $q$ from \eqref{2q} and \eqref{q} in order to obtain an explicit function $\Delta_2(\Delta)$. The analytic expression, however, is not illuminating, so we only show the 
 asymptotic behavior. In the large $c$ limit, we have
\begin{align}
\Delta_2 = 2\Delta + 3\Delta^3 c^{-2} + \cdots 
\end{align}
and in the small $c$ limit, we have
\begin{align}
\Delta_2 = 2\sqrt{2} \Delta -\frac{6242\ 2^{5/6} ({c^2}\Delta)^{1/3}}{6561\ 3^{29/36}} + \cdots 
\end{align}
In between, we have a smooth extrapolation of the two straight lines $\Delta_2 = 2\Delta$ and $\Delta = 2\sqrt{2}\Delta$ with no significant features. The change of the slope occurs when the size of the black hole is comparable with the AdS radius $L$. We will show the numerical plot in the next section.

\section{Conformal bootstrap bound}
From holography, we can map the spectrum of the Reissner-Nordstr\"om black hole in $D = d+1$ dimensional AdS space-time into the conformal data of a dual conformal field theory in $d$ dimensions. Here, we study the bound of the conformal data of a generic conformal field theory with a $U(1)$ global symmetry in three dimensions. The $U(1)$ global symmetry corresponds to the existence of the $U(1)$ gauge field that supports the Reissner-Nordstr\"om black hole. 

Consider a four-point function $\langle \Phi_q \Phi_q \Phi_{-q} \Phi_{-q} \rangle$ of spinless operators with charge $q$ and conformal dimension $\Delta$. We will assume  $\Phi_q$ as the operator that corresponds to the charge $q$ extremal Reissner-Nordstr\"om black hole (or more precisely its micro state). We use the operator product expansion and decompose the four-point functions into conformal blocks. Then, the crossing symmetry gives conformal bootstrap equations (see e.g. \cite{Poland:2018epd} for a review) with the $U(1)$ global symmetry \cite{Rattazzi:2010yc}\cite{Vichi:2011ux}\cite{Poland:2011ey}\cite{Kos:2013tga} 
\begin{align}
0 = \sum_{S^+} \lambda^2_{S^+} \left(
    \begin{array}{c}
      0 \\
      F  \\
      H 
    \end{array}
  \right) + \sum_{T^+} \lambda^2_{T^+}  \left(
    \begin{array}{c}
      F \\
      0  \\
      -2 H
    \end{array}
  \right) + \sum_{A^-} \lambda^2_{A^-}  \left(
    \begin{array}{c}
      -F \\
       F  \\
      -H 
    \end{array}
  \right)  \label{cross}
\end{align} 
where $(\pm)$ denotes the even $(+)$ or odd $(-)$ spin contributions. We have used the convention
\begin{align}
F & = v^{\Delta_{\Phi}} g_{\Delta_O,l}(u,v) - u^{\Delta_{\Phi}} g_{\Delta_O,l}(v,u) \cr
H & = v^{\Delta_{\Phi}} g_{\Delta_O,l}(u,v) + u^{\Delta_{\Phi}} g_{\Delta_O,l}(v,u)
\end{align}
with the conformal block $g_{\Delta_O,l}$ being normalized as in \cite{Hogervorst:2013kva} (which can be explicitly found in \cite{Dolan:2003hv}). The unitarity assumes $\lambda^2_O \ge 0$ and $\Delta_O \ge d-2+l$. Here $d=3$.

\begin{figure}[htbp]
	\begin{center}
		\includegraphics[width=12.0cm,clip, bb= 0 350 640 750]{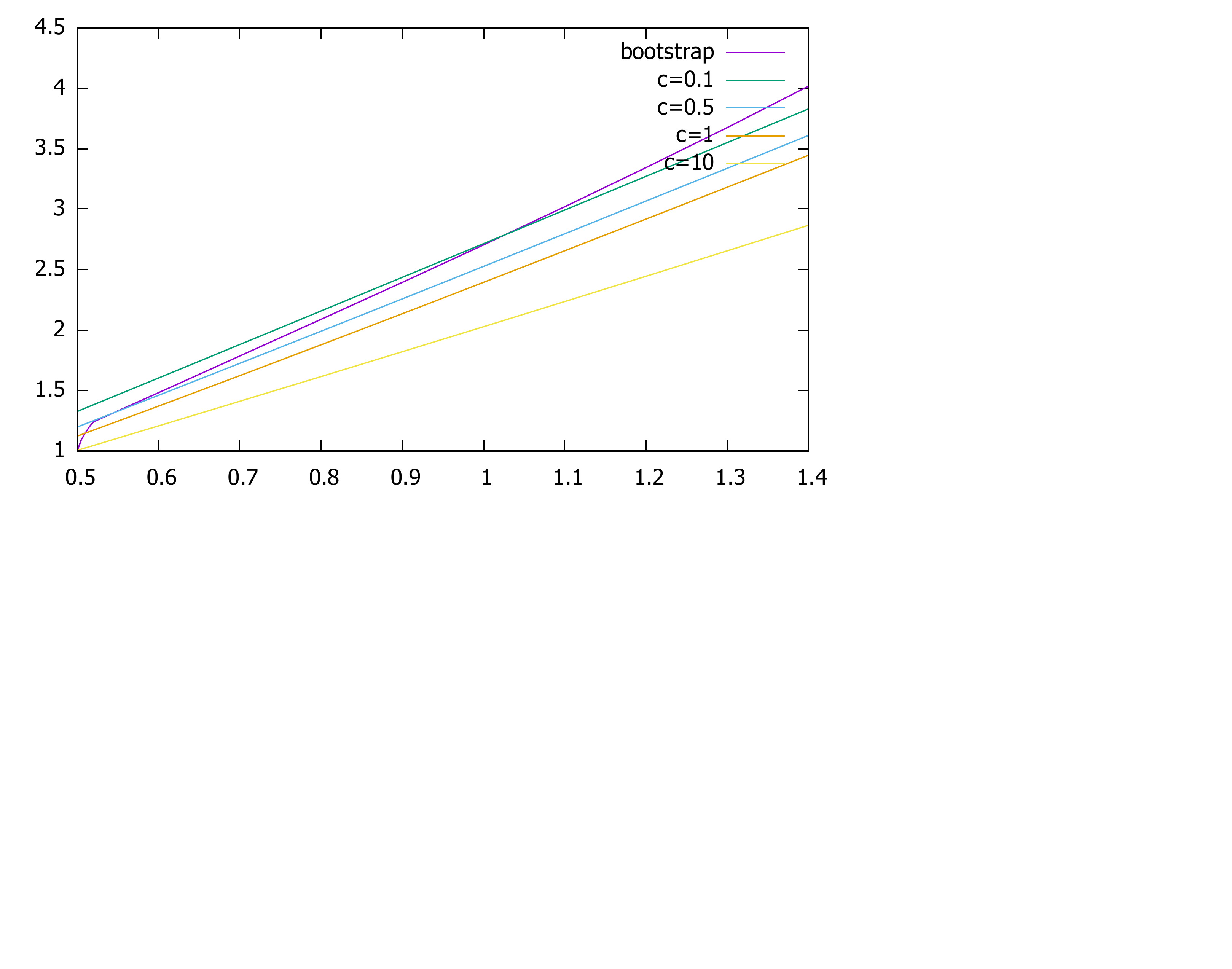}
	\end{center}
	\caption{Unitarity bound for $\Delta_{2}$ in three-dimensional $O(2)$ symmetric conformal field theory as a function of $\Delta$.}
	\label{fig:1}
\end{figure}

Under the unitarity assumption, the conformal bootstrap equations \eqref{cross} become a semi-definite program and it can be numerically analyzed by using existing software \cite{Simmons-Duffin:2015qma}\cite{cboot}. In Figure \ref{fig:1}, we show the bound of the conformal dimension $\Delta_2$ of the twice charge operator, which is the lowest energy state in $T^+$ sector, as a function of $\Delta$. The bound means that there must exist an operator whose conformal dimension is lower than the curve.
 We also compare the bound with the extremal Reissner-Norstr\"om black hole spectrum.

When the Reissner-Norstr\"om spectrum is above the conformal bootstrap bound, such an energy spectrum is inconsistent with the holographic interpretation with no other lower energy states. The smaller the $c$ is, the larger the excluded region. This is expected because small $c$ means a large quantum gravity correction if any. We note that whatever the value of $c$ is, the ``smallest"  Reissner-Norstr\"om  black hole with $\Delta=\frac{1}{2}$ is inconsistent with the conformal bootstrap bound. 

As we have observed in \cite{Nakayama:2019jvm}, it is not obvious if the numerical bound is optimal for larger $\Delta$. It is possible with advanced knowledge of the conformal bootstrap analysis, we may further constrain the spectrum in the larger $\Delta$ regime. See e.g. \cite{Jafferis:2017zna}. We also note that we did not use any information from the central charge in the conformal bootstrap analysis. Extra input from the central charge may give a stronger bound than what we have presented in this section.

\section{Discussions: how to resolve the inconsistency?}
We have seen that the spectrum of the extremal Reissner-Nordstr\"om black hole  is inconsistent with the conformal bootstrap bound in the regime of small $\Delta$. Now we would like to discuss how this inconsistency is resolved in a theory of quantum gravity. 

The first option is that the energy spectrum of the Reissner-Nordstr\"om black hole is modified by quantum gravity corrections. For example, we may simply forbid the existence of the Reissner-Nordstr\"om black hole that violates the conformal bootstrap bound. After all, the unitarity of the AdS algebra demands $\Delta \ge \frac{1}{2}$, and the energy of the black hole cannot be lower than this although the classical black hole solution itself does not appear pathological.\footnote{A pathology comes from the action of the momentum operator that leads to a negative norm state in the dual conformal field theory. The action of the momentum operator is an asymptotic symmetry of the AdS space-time, so the classical solution would not immediately see the pathological behavior.}

Alternatively, one may modify the spectrum of the Reissner-Nordstr\"om black hole as a function of $q$. There are various possibilities here, but one simple scenario is to add a constant to the energy
\begin{align}
\tilde{\Delta}(q) =  c \frac{\sqrt{-1+\sqrt{1+12q^2}}(2+\sqrt{1+12q^2})}{3\sqrt{6}}  + \delta_0 \ . 
\end{align}
The constant shift $\delta_0$ may be associated with the one-loop effects in quantum gravity. This is also motivated from the large $q$ expansion of conformal dimensions in $U(1)$ symmetric conformal field theory, where the existence of the constant term $\delta_0$ is universally predicted \cite{Hellerman:2015nra}.

With this simple shift, $\tilde{\Delta}_2 = \tilde{\Delta}(2q)$ as a function of $\tilde{\Delta}$ is modified as
\begin{align}
\tilde{\Delta}_2 = -\delta_0 +2\tilde{\Delta} - \frac{9}{c^2} \delta \tilde{\Delta}^2 + \frac{3}{c^2} \tilde{\Delta}^3 + \cdots \ .
\end{align}
in the small $\tilde{\Delta}$ limit. In order for this spectrum to be consistent with the conformal bootstrap bound, we need to fix $\delta_0$ as 
\begin{align}
\delta_0 = \frac{3}{8c^2} + \cdots 
\end{align}
in the large $c$ limit. The shift essentially moves the curve in  Figure \ref{fig:1} so that  $(\tilde{\Delta}, \tilde{\Delta}_2) = (1/2,1)$ is now on it, and we see that the entire curve is inside the conformal bootstrap bound.

The other possibility is that the weak gravity conjecture holds.\footnote{We emphasize that the weak gravity conjecture in the AdS space-time is non-trivial. See \cite{Huang:2006hc}\cite{Li:2015rfa}\cite{Harlow:2015lma}\cite{Benjamin:2016fhe}\cite{Heidenreich:2016aqi}\cite{Montero:2016tif}\cite{Conlon:2018vov}\cite{Montero:2018fns}\cite{Alday:2019qrf}\cite{Cremonini:2019wdk}\cite{Agarwal:2019crm} for various approaches in relation to holography.} We have assumed that the lowest energy state with a given charge is given by the extremal Reissner-Nordstr\"om black hole, but if this were not the case, the inconsistency with the conformal bootstrap bound could be gone simply because the operator that satisfies the conformal bootstrap bound would be different from the Reissner-Nordstr\"om black hole.  In some cases, the theory includes light charged scalar fields and the hairy black hole would be a candidate of the lower energy states. In many supersymmetric situations, this is often the case and the resultant hairy black hole saturates the BPS bound and therefore $\Delta_2 = 2\Delta$. However, we should rethink if the stability of the hairy black hole is consistent with the quantum gravity constraint. What would be fundamental difference between the stability of the extremal Reissner-Nordstr\"om black hole and the extremal hairy black hole? Perhaps, the hairy black holes are more difficult to be distinguished with the elementary particles, but we would like to establish this picture more firmly.


\section*{Acknowledgements}
The author would like to thank D.~Orlando and S.~Reffert for their visit to Rikkyo university and stimulating discussions.
This work is in part supported by JSPS KAKENHI Grant Number 17K14301.

\end{document}